\documentclass[a4paper]{article}

\usepackage{INTERSPEECH2021}
\usepackage{xcolor}
\usepackage{soul}

\usepackage[noadjust]{cite}

\title{Automatic Analysis of the Emotional Content of Speech in Daylong Child-Centered Recordings from a Neonatal Intensive Care Unit}
\name{Einari Vaaras$^1$, Sari Ahlqvist-Björkroth$^2$, Konstantinos Drossos$^1$, Okko Räsänen$^{1,3}$}
\address{
  $^1$Unit of Computing Sciences, Tampere University, Finland\\
  $^2$Department of Clinical Medicine, University of Turku, Finland\\
  $^3$Department of Signal Processing and Acoustics, Aalto University, Finland}
\email{einari.vaaras@tuni.fi, sarahl@utu.fi, konstantinos.drossos@tuni.fi, okko.rasanen@tuni.fi}

\begin{document}

\maketitle
\begin{abstract}

Researchers have recently started to study how the emotional speech heard by young infants can affect their developmental outcomes. 
As a part of this research, hundreds of hours of daylong recordings from preterm infants' audio environments were collected from two hospitals in Finland and Estonia 
in the context of so-called APPLE study. In order to analyze the emotional content of speech in such a massive dataset, an automatic speech emotion recognition (SER) system is required. However, there are no emotion labels or existing in-domain SER systems to be used for this purpose. In this paper, we introduce this initially unannotated large-scale real-world audio dataset and describe the development of a functional SER system for the Finnish subset of the data. We explore the effectiveness of alternative state-of-the-art techniques to deploy a SER system to a new domain, comparing cross-corpus generalization, WGAN-based domain adaptation, and active learning in the task. As a result, we show that the best-performing models are able to achieve a classification performance of 73.4\% unweighted average recall (UAR) and 73.2\% UAR for a binary classification for valence and arousal, respectively. The results also show that active learning achieves the most consistent performance compared to the two alternatives. 
  
  
\end{abstract}

\section{Introduction}

In speech emotion recognition (SER), the task is to
recognize emotional states of speakers from speech signals \cite{computational_paralinguistics,batliner_recognition_of_emotions_in_speech}. 
One potential 
application of SER is the study of babies' auditory environments, where the early emotional experiences of babies, including affective speech, can impact their later cognitive development. In order to study this relationship, \textit{Auditory environment by Parents of
Preterm infant; Language development and Eye-movements} (APPLE) study has collected a large audio corpus of child-centered daylong audio recordings from neonatal intensive care units (NICUs), recorded in Turku University Hospital, Finland, and Tallinn Children's Hospital, Estonia \cite{eva_paperi}. In order to analyze the emotional contents of speech in the recordings, a functional SER system for this new domain is required.

The purpose of the present study is to develop such a system to analyze these (initially unannotated) hospital-environment audio recordings for their emotional speech content.
The absence of 
in-domain annotations and massive scale of the data raises the question of how to most effectively deploy a SER system for this real-world large-scale dataset. 


In principle, \textit{cross-corpus generalization} (CCG) is the most straightforward strategy to deploy SER for an unlabeled dataset, but can suffer from domain mismatch. In fact, \cite{schuller_cross_corpus_ser} have shown through extensive multi-corpus and multilingual experiments that reliable CCG-based SER was only feasible with certain corpora and emotional classes, highlighting many issues with cross-domain SER model generalization to out-of-domain data (but see also, e.g., \cite{ser_f_similarity} for a potential remedy).   
In order to tackle the issue of domain mismatch, different \textit{domain adaptation} (DA) methods have been utilized in SER. For instance, Deng et al.
\cite{ser_semisupervised_autoencoders} extended an unsupervised deep denoising autoencoder (AE) by combining
it with a supervised learning objective to create a semi-supervised DA method for SER. 
Another approach in
\cite{ser_domain_adversarial} used an unsupervised deep neural network (DNN)-based adversarial DA approach for SER. The
method
learns a domain-invariant feature representation between labeled source data and unlabeled 
target-domain data while maintaining a good performance on the primary SER task. A number of other DA methods for SER have been proposed as well (e.g.,
[\citen{ser_da_cross_lingual,ser_universum_autoencoder,ser_cross_lingual}]).

\textit{Active learning} (AL) is another strategy and has been successfully applied to SER as well. Zhao and Ma 
\cite{ser_al_random_fields} presented an iterative AL algorithm, which utilizes conditional random fields, to determine the level of uncertainty for each unlabeled sample. The most uncertain samples were then selected for human annotation. Another study \cite{ser_dnn_al} examined different AL methods based on uncertainty and diversity maximization in a simulation setup with DNN classifiers. The work showed that the tested AL methods outperformed random sampling-based methods with a constrained labeling budget.


Only a few SER studies have been conducted on large-scale datasets. Jia et al. \cite{ser_largescale_internet} studied DNN-based SER with 
a massive 7-million-utterance internet voice corpus. 
They pretrained their novel DNN-based models with 90,000 unlabeled utterances, and fine-tuned and evaluated them on 3,000 randomly selected manually annotated utterances from the same dataset.
Fan et al. \cite{lssed} presented a 
SER dataset with 
a total duration of over 200 hours. They proposed a novel SER model
containing pyramid convolutions which outperformed other models that were tested on the dataset. Additionally, they
showed that existing models are prone to overfit to small-scale datasets, which limits the ability of these models
to generalize for real-life data.

However, CCG, AL, and DA have rarely been compared to each other directly. Moreover, most of the existing work has been conducted using studio, telephone, or internet speech data. Therefore, our present daylong audio dataset from a hospital context, together with its practical significance, provides an excellent test bench to compare strategies for SER system development in a novel domain with challenging real-world speech data. More specifically, by using the Finnish subset of the data, we compare CCG and state-of-the-art DA and AL in the task to study their feasibility and SER performance in practice. 



\section{Methods}

\subsection{Medoid-based active learning} \label{sec_mal}

Zhao et al. \cite{tuni_mal} presented an AL method called medoid-based active learning (MAL) to effectively utilize a small number of annotations, which serves as the foundation of the AL method used in our experiments. The algorithm can be divided into three subsequent parts: 1) obtaining a distance matrix that contains the pairwise distances between all samples in the dataset, 2) performing $k$-medoids clustering using the distance matrix, and 3) starting from the largest cluster, querying human annotations for the medoids in a descending cluster size order.

The distance metric used in the present experiments was selected based on pilot
experiments with MAL using existing SER datasets. A 600-dimensional utterance-level log-mel feature representation (see Section \ref{sec_features}) was first used as the initial feature representation of each sample in a dataset. These features were then compressed into a 32-dimensional latent representation using a DNN-based AE with six layers. Pearson distances $d_P$ \cite{pearson_distance} between the bottleneck features were then used to define the affinity matrix $A$ across all the samples.
Next, $k$-medoids clustering was applied to the data. First, one
sample was randomly selected as the member of a set $S$, followed by an addition of $k-1$ more samples as centroids using the \textit{farthest-first traversal} algorithm. 
Here, the distance from a sample, $\bm{a}$, to the set $S$ was defined as
\vspace{-6 pt}

\begin{equation}
\label{eq_farthest_first}
    d_P(\bm{a},S) = \underset{\bm{b}\in S}{\textrm{min}} \ d_P(\bm{a},\bm{b}) \ .
\vspace{-2 pt}
\end{equation}
The samples in $S$ were then used as the initial medoids for a $k$-medoids clustering algorithm (see e.g. \cite{k_medoids_algorithm} for an overview) to assign each sample in the dataset into one of the clusters.

In the final stage, the clusters were sorted in a descending order based on the number of samples in each cluster, and their medoids were presented to human annotators for labeling. In the experiments, we studied the use of these labels in two different ways: i) assigning each sample in a cluster with the annotated medoid label (as in \cite{tuni_mal}; here referred to as \textit{``cluster labels"}), or ii) only using the medoid samples as labeled data for classifier training, which was not studied in the original MAL paper \cite{tuni_mal}.
Based on pilot experiments on other datasets, $k$ was set to $\frac{N}{3}$, where $N$ is the number of samples in a corpus.

\subsection{Wasserstein distance-based domain adaptation}

\begin{figure}[t]
    \centering
    \vspace{-3 pt}
    \includegraphics[width=0.3\textwidth]{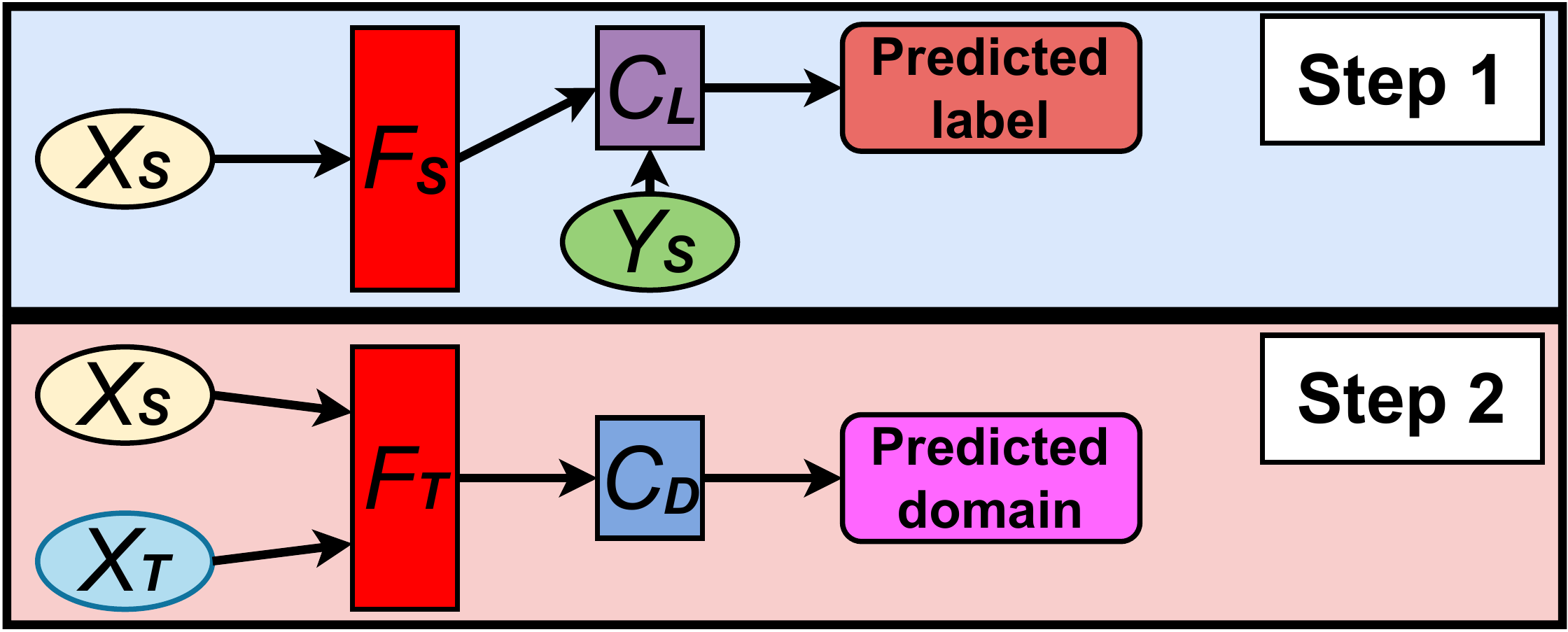}
    \vspace{-7 pt}
    \caption{The two-step the adaptation process of WDA. First, $F_{S}$ and $C_L$ are trained to classify source corpus samples into emotion categories. In the second step, 
    $F_{S}$ is adapted into $F_{T}$ using a domain discriminator $C_{D}$ with an adversarial loss.}
    \label{fig:undaw_two_stages}
    \vspace{-18 pt}
\end{figure}

The present DA approach was based on the Wasserstein distance-based domain adaptation (WDA) method proposed in \cite{tuni_adversarial_domain_adaptation}. In WDA, a neural network (NN) classifier, aka the \textit{source model} $M$, is adapted to a target corpus, $D_T$, by using labeled data from source domain corpus/corpora, $D_S$.
The source model $M$ consists of two parts, a feature extractor, $F_S$, and a label classifier,
$C_L$. The adaptation process of WDA involves two stages, which are demonstrated in Fig.
\ref{fig:undaw_two_stages}. 

The first stage (Fig. \ref{fig:undaw_two_stages}, top) consists of training $M$ using samples $X_S$ and their labels $Y_S$ from $D_S$ to obtain a trained $F_S$. This is done using binary cross-entropy
\cite{tuni_adversarial_domain_adaptation} as the loss:

\vspace{-12 pt}
\begin{equation} \label{wda_binary_cross_entropy}
    L_M(\bm{x},\bm{y}) = -\sum_{(\bm{x},\bm{y})\in(X_S, Y_S)}{{\bm{y}^T} log_{10}(C_L(F(\bm{x})))} .
\end{equation}
\vspace{-8 pt}

In the second stage (Fig. \ref{fig:undaw_two_stages}, bottom), $F_S$ is adapted to $D_T$ to obtain an adapted feature extractor, $F_T$, by minimizing the Wasserstein-1 distance $W_d$ between the distributions of $D_S$ and $D_T$ using an adversarial training process. Following a WGAN framework \cite{wgan_original}, $F_S$ is adapted into $F_T$ by finding a common feature representation for $D_S$ and $D_T$ by iteratively minimizing the two losses:
%
\begin{equation} \label{wda_adaptation_loss_1}
    L_{C_D}(\bm{x},\bm{z}) = \sum_{\bm{x} \in X_S}{C_D(F_S(\bm{x}))} - \sum_{\bm{z} \in X_T}{C_D(F_T(\bm{z}))}
\end{equation}
\vspace{-6 pt}
\begin{equation} \label{wda_adaptation_loss_2}
    L_{F_T}(\bm{x},\bm{y},\bm{z}) = \sum_{\bm{z}\in X_T}{C_D(F_T(\bm{z}))} + L_M(\bm{x},\bm{y}) \ ,
\end{equation}
where $C_D$ is the domain discriminator and $X_T$ are the target corpus samples. The parameters for $C_D$ and $F_T$ are updated in turns, where
Eqs. \ref{wda_adaptation_loss_1} and \ref{wda_adaptation_loss_2} are the loss functions for updating the
parameters of $C_D$ and $F_T$, respectively. The output features of $F_T$ are the input features for $C_D$.
Additionally, the parameters of $F_S$ serve as the initial parameters of $F_T$. As pointed out in \cite{tuni_adversarial_domain_adaptation}, the minimization of Eqs. \ref{wda_adaptation_loss_1} and \ref{wda_adaptation_loss_2} is shown to minimize $W_d$ between the distributions of $D_S$ and $D_T$. For a detailed formulation of the WDA algorithm, see Algorithm 1 in \cite{tuni_adversarial_domain_adaptation}.

\subsection{Cross-corpus generalization}

As our baseline approach, we use CCG with different source corpora and their combinations. Labels of each corpus are first mapped to a
common emotion category space, followed by a standard supervised classifier training (see Section
\ref{sec_conducted_experiments}).

\section{Data}

\subsection{NICU-A} \label{sec:nicu}

The FinEst NICU Audioset (NICU-A) was collected in the APPLE study, and is the primary audio material for which our SER system was aimed to be deployed on. We use the Finnish subset of the dataset, which was recorded at the NICU of Turku University Hospital using LENA-recorders (https://www.lena.org/) placed at the bedside of preterm babies (average age approx. 33 gestational weeks) in intensive care. The data consists of 43 x 16-hour recordings from different participating families (a total of 688 h of audio). The recordings were carried out in relatively calm single family rooms of the NICU, where only the baby, visiting parents (primary talkers), and, occasionally, nurses and doctors carrying out healthcare routines were present.

Broad-class diarization of LENA software \cite{LENAXu} was used to split each 16-h recording into utterance-sized segments, and to assign a speaker tag (male, female, key child, other child) to the utterances. Based on the validity study reported for the same data in \cite{siirila_kati_gradu}, adult speech from ``male/female adult" ``near" and ``far" -categories were included in the analyses to capture caregiver speech (but see \cite{lena_evaluation} for general guidelines with LENA ``far" data). Utterances shorter than 600 ms were discarded from further analysis. This resulted in a total of 129,007 utterances with an average length of 1.57 s (approx. 56 h of speech).

Eight families were carefully selected as the test data and 35 as the training data based on the representativeness of both data sets in terms of
covariates such as child health, parental presence etc. After pre-processing the data of NICU-A, both the training and test sets were partially annotated.


For the training data, samples were selected for annotation using MAL, as described in Section \ref{al_exp}. Two
annotators performed labeling for distinct subsets of the data,
except for the first 200 samples that were annotated by both to measure inter-rater agreement rates. Each sample was annotated in two dimensions:
in terms of binary arousal (high/low) and in terms of ternary valence (negative, neutral, positive). The two dimensions were annotated in a random order for each sample. A sample could also be labeled as erroneous, if the samples were corrupted by noise, had overlapping speakers, had very short speech fragments, or did not contain speech at all.

For the test data, gold standard (GS) annotations were obtained from three speech/clinical experts for a randomly selected subset of samples from the test set. All GS samples were independently annotated for their arousal and valence by all three annotators, followed by majority voting of labels. Samples without majority labels were removed from the test set. GS annotators had access to 10 s of the preceding audio context of each sample to better understand the communicative context. 

After removing the erroneous files, the training and test sets had 5198 and 345 labeled samples, respectively.
Training data inter-annotator agreement rates in terms of kappa scores were 0.78 for valence and 0.64 for arousal. For the GS data, the kappa scores were 0.48 for valence and 0.28 for arousal. The difference between the training and testing agreement rates is explained due to the use of MAL in the selection of the training samples, where the first 200 samples annotated by both annotators were also the most acoustically distinct samples in the training data. The finding also demonstrates the inherent difficulty in annotating a random sample of real-world speech for emotional content.

The `neutral' and `negative' classes for valence were merged for NICU-A, bearing in mind that the APPLE study was primarily interested in the proportion of positive valence over other speech. As a result, training sample counts were 1509 for positive and 3689 for neutral valence, and 3165 and 2033 for high and low arousal, respectively. The corresponding test set counts were 120 (positive) and 225 (neutral) for valence, and 89 (high) and 256 (low) for arousal. 

\subsection{Other corpora for CCG and DA experiments} \label{sec:other_corpora}

In addition to NICU-A, four existing SER corpora (referred to as \textit{source corpora}) were used in the CCG and DA experiments:

\textit{The Berlin Emotional Speech Database} (EMO-DB) \cite{emodb} is a widely used corpus and consists of 535 spoken utterances in German from 10 professional actors with seven emotional labels: anger, boredom, disgust, fear, joy, neutral, and sadness.

\textit{eNTERFACE} \cite{enterface} is an audiovisual database consisting of 1287 video samples in English from 42 test subjects from 14 nationalities in six categories: anger, disgust, fear, joy, sadness, and surprise. Only the audio tracks were used in this study.

\textit{The Finnish Emotional Speech Corpus} (FESC) \cite{finnish_corpus} consists of nine professional actors portraying emotions of five different categories: neutral, sadness, joy, anger, and tenderness. These portrayals were split into 4254 utterances based on long silences as defined by an energy threshold \cite{einari_dippa}.
    
\textit{The Ryerson Audio-Visual Database of Emotional Speech and Song} (RAVDESS) \cite{ravdess} is a multimodal database including a total of 7356 recordings from 24 professional actors, out of which 1440 speech-only recordings were used in the present study. Eight different emotional labels were included: neutral, calm, happy, sad, angry, fearful, surprise, and disgust.

\section{Experimental setup}

\subsection{Features} \label{sec_features}

Log-mel, GeMAPS, and eGeMAPS \cite{gemaps} features  were used in the CCG and AL experiments. For the DA experiments, only log-mel
features were used due to their superior performance in pilot experiments. For the log-mel features, 40 mel
filters were used with a Hann window using a 30-ms window size and 10-ms shifts. To get constant-dimensional
utterance feature representations, seven functionals (the first four moments, min, max,
and range) were taken from the time series of the log-mel features. In addition, four functionals (the first four
moments) were applied to first and second order delta features. This resulted in a 600-dimensional feature vector for the log-mel features.
The 62- and 88-dimensional GeMAPS and eGeMAPS features were extracted using the openSMILE toolkit
\cite{opensmile}. The features for each corpus were z-score normalized at the corpus level.

\subsection{Conducted experiments} \label{sec_conducted_experiments}

For the source corpora, the emotional labels were mapped into the quarters of the valence-arousal plane following \cite{schuller_cross_corpus_ser}, with the exception of merging `neutral' and 'negative' valence to `neutral' in order to better correspond to the labels of NICU-A. The emotional mapping of \cite{schuller_cross_corpus_ser} has been used in multiple SER studies (e.g. [\citen{ser_da_cross_lingual,ser_cross_lingual,ser_unite_or_vote,ser_unsupervised_learning_cross_corpus}]). All classification tests were conducted on the NICU-A GS data. We use the unweighted average recall (UAR \%) as the primary evaluation measure.

\subsubsection{Active Learning Experiments} \label{al_exp}

In the AL experiments, MAL was performed for the full unlabeled training set of NICU-A (101,813 samples). To compress the log-mel features of the training set into a latent representation, an AE network was used. The training and validation data for the AE were based on a random split of the training set using a 
ratio of 80:20 utterances. The encoder of the AE consisted of three fully-connected (FC) ELU \cite{elu_original} layers of 512, 512, and 32 units, and the decoder of two 512-unit ELU layers and a linear reconstruction layer. The first two AE layers had a dropout of 0.1. The model was trained using MSE loss, Adam \cite{adam_original} optimizer (\textit{lr} = $10^{-4}$), batch size of 1024, and early stopping with a patience of 300. The best model according to the validation loss was then used to compress the data to 32 dimensions. Then, MAL was performed for each of the 35 training set families separately and the data were sent for annotation (Section \ref{sec:nicu}).

The annotated samples were then used for training a support vector machine (SVM) with an RBF kernel. Each sample was weighted inversely proportional to its class frequency to counter class distribution imbalances. Optimal SVM hyperparameters were selected for each feature type and both classification tasks individually based on a grid search using 5-fold cross-validation over the training data. Then, the SVM was trained on the full training data using these hyperparameters and tested on the GS data. The process was performed separately for the labeled training set of 5,198 samples and for the extended training set of 33,979 samples using the cluster labels from MAL. 

\subsubsection{Cross-corpus Generalization Experiments} \label{ccg_exp}

For the CCG experiments, two settings were explored: 1-to-1 and 4-to-1 CCG. In the 1-to-1
setting, each of the source corpora was used individually as the training set. In the 4-to-1
setting, all four source corpora were used for SVM training with similar specifications as with the AL experiments. 

\subsubsection{Domain Adaptation Experiments}

\begin{table}[t]
    \vspace{-10 pt}
    \centering
    \caption{
    UAR (\%) performance scores for alternative approaches on the target data. For AL and CCG, log-mel (log-m), GeMAPS (Ge), and eGeMAPS (eGe) features are compared. For DA, the unsupervised (US) and semi-supervised (S-S) variant of WDA is compared. The highest accuracies are \textbf{bolded}.} 
    \vspace{-8 pt}
    \includegraphics[width=0.47\textwidth]{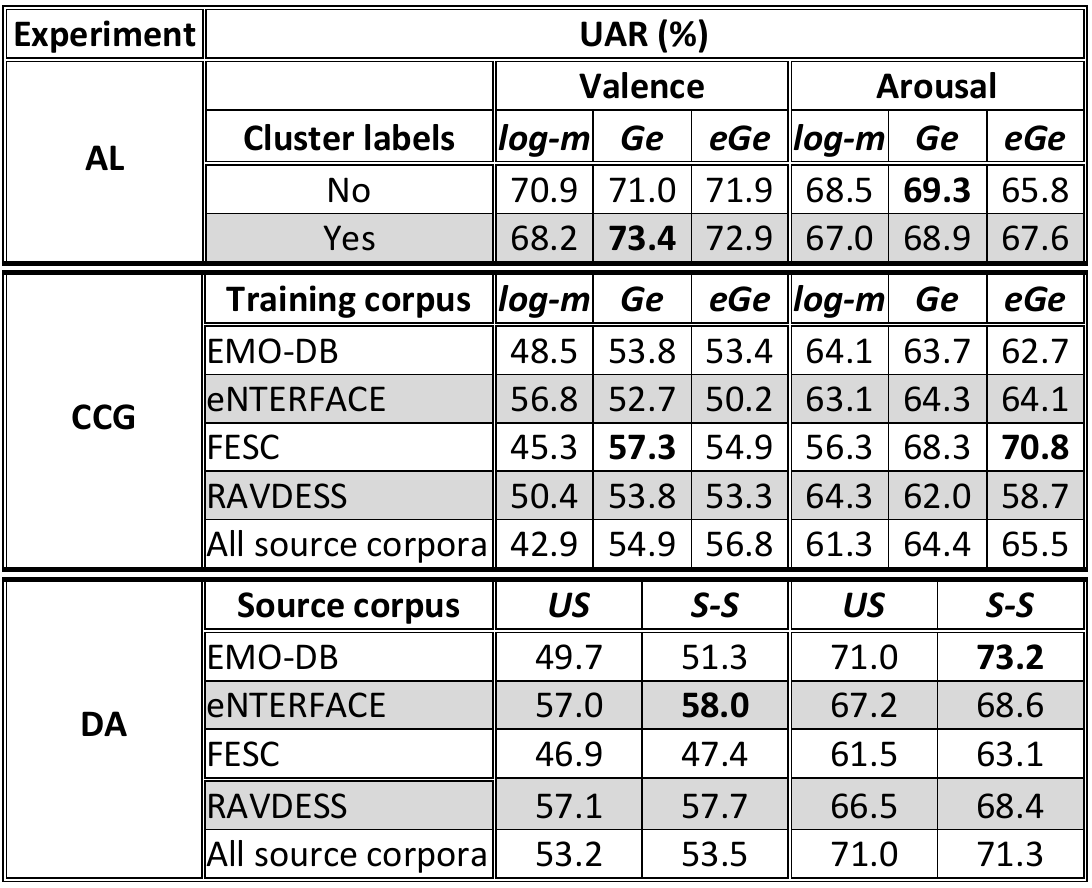}
    \vspace{-26 pt}
    \label{table_utu_results}
\end{table}

For the DA-based experiments, 1-to-1 and 4-to-1 adaptation conditions were examined with the same source corpora as in CCG. All DA experiments were conducted separately for valence and arousal.
In the 1-to-1 settings, each source corpus was randomly split into a training and test set in a ratio of 85:15. For the
4-to-1 setting, the training and test sets were the combination of the respective corpus-specific
splits. For the first stage of the adaptation process, the training set of each source corpus was
used to train $M$ by using the Adam optimizer (\textit{lr} = $10^{-4}$), early stopping with a patience of 100 based on test set accuracy, and batch size of 256. 
The log-mel features were used as the input features for $F$, consisting of three FC layers of 512, 512, and 256 units, each followed by batch normalization. The first two layers had LReLU \cite{leakyrelu_original} nonlinearities and a dropout of 0.4. $C_L$ was an NN consisting of three FC layers of 256, 256, and 2 units. The first two layers had LReLU nonlinearities and a dropout of 0.3. The last layer was followed by a softmax function. For each variant of the source data, a separate $M$ was trained for both valence and arousal.

For the second stage of the adaptation process, the full unlabeled data from the source corpus/corpora and the unlabeled training samples of NICU-A were used for training. Following \cite{tuni_adversarial_domain_adaptation}, the unsupervised variant of WDA was trained until the first
term in Eq. \ref{wda_adaptation_loss_2} was saturated. For the semi-supervised variant, the labeled training set of NICU-A was used to determine the model accuracy after each epoch, and the model with the highest accuracy was selected for testing. This set was also used to find optimal hyperparameters. $C_D$ consisted of four FC layers of 512, 512, 256, and 1 units. The first three layers were followed by ReLU nonlinearities. The parameters of $C_D$ and $F_T$ were updated with the RMSProp \cite{rmsprop_original} and Adam optimizers, respectively. In the 1-to-1 settings, \textit{lr} = $5\cdot10^{-5}$ was used, except with FESC for valence and with RAVDESS for arousal, where \textit{lr} = $7\cdot10^{-5}$. For the 4-to-1 settings, \textit{lr} = $7\cdot10^{-5}$ was used for valence and \textit{lr} = $6\cdot10^{-5}$ for arousal. The performance of the adapted model was then tested on the GS data.

All the DA and AL parameters were based on extensive piloting with leave-one-corpus-out simulations using the source corpora, and before any NICU-A data had been labeled. 

\vspace{-1 pt}
\section{Results}
The main results are presented in Table \ref{table_utu_results}. They show that AL (top rows) is the most consistent performer across the studied conditions, even though somewhat better arousal results are obtained by particular configurations of CCG and DA. The best DA-based model adaptation achieves 73.2\% UAR on arousal, outperforming all other methods by a clear margin. However, adaptation from other corpora does not always work that well. In addition, CCG and DA have problems with valence classification on data from the new domain. The DA results (Table \ref{table_utu_results}, bottom) are on average higher than the results of CCG, even though the WDA method does not provide a major improvement over CCG on valence. The semi-supervised variant of WDA is also consistently better than the unsupervised variant. 
The comparison of using either cluster or medoid labels for AL provides somewhat mixed results, depending on the exact condition.  

In terms of features, the GeMAPS and eGeMAPS feature sets outperformed the log-mel features on valence with CCG. For CCG and arousal, the best-performing features varied largely between different training corpora, and the matching Finnish language FESC is a substantially better source for NICU-A than the others, reaching 70.8\% UAR with eGeMAPS features. In the AL experiments (Table \ref{table_utu_results}, top), the eGeMAPS and GeMAPS features achieved the best mean classification accuracy for valence and arousal, respectively.

\begin{figure}[t]
    \vspace{-12 pt}
    \centering
    {{\includegraphics[width=0.47\textwidth]{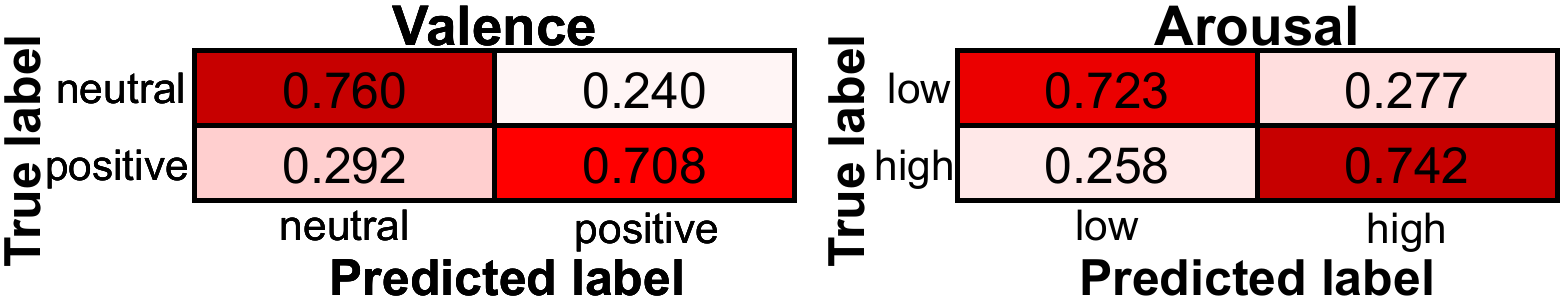} }}
    \vspace{-18 pt}
    \caption{Normalized confusion matrices for valence (left) and arousal (right) using the best models. Valence = SVM + GeMAPS + cluster labels from MAL (73.4\% UAR). Arousal = NN + WDA using EMO-DB as the source corpus (73.2\% UAR).} 
    \vspace{-13 pt}
    \label{fig:confusion_matrices}
\end{figure}


The confusion matrices for the best-performing models (Fig. \ref{fig:confusion_matrices}) indicate that these models do not systematically favor one label over the other when performing predictions.

\section{Conclusions}

In the present paper, we developed a SER system for large-scale analysis of emotional content of speech in initially unannotated real-life child-centered audio recordings from a NICU. CCG, AL, and DA were
compared as alternatives for deploying a SER system for this novel dataset from scratch. Our results show that WGAN-based DA outperformed the baseline CCG approach, verifying its usefulness in the absence of any data labels. However, with a very moderate human labeling resource available, $k$-medoids based AL was superior compared to CCG and DA in valence classification and relatively competitive for arousal as well. However, when classifying arousal, DA resulted in slightly better results than AL. Overall, the results demonstrate that the earlier proposed MAL \cite{tuni_mal} and WDA \cite{tuni_adversarial_domain_adaptation} methods are also applicable to practical SER scenarios. The results also show that emotion analysis for LENA-based daylong audio recordings is possible with an accuracy comparable to those reported in earlier literature
(e.g., 58.1\% for valence and 66.8\% for arousal across the multi-corpus tests in \cite{ser_unsupervised_learning_cross_corpus}).

\vspace{-1 pt}
\section{Acknowledgements}
OR was funded by Academy of Finland grant no. 314602 and KD partially by EU Horizon-2020 grant no. 957337 MARVEL. SAB was funded by Academy of Finland grant no. 332962. The authors thank the APPLE consortium for the help in the project.

\bibliographystyle{IEEEtran}

\bibliography{mybib}

\begin{thebibliography}{10}
\providecommand{\url}[1]{#1}
\csname url@samestyle\endcsname
\providecommand{\newblock}{\relax}
\providecommand{\bibinfo}[2]{#2}
\providecommand{\BIBentrySTDinterwordspacing}{\spaceskip=0pt\relax}
\providecommand{\BIBentryALTinterwordstretchfactor}{4}
\providecommand{\BIBentryALTinterwordspacing}{\spaceskip=\fontdimen2\font plus
\BIBentryALTinterwordstretchfactor\fontdimen3\font minus
  \fontdimen4\font\relax}
\providecommand{\BIBforeignlanguage}[2]{{%
\expandafter\ifx\csname l@#1\endcsname\relax
\typeout{** WARNING: IEEEtran.bst: No hyphenation pattern has been}%
\typeout{** loaded for the language `#1'. Using the pattern for}%
\typeout{** the default language instead.}%
\else
\language=\csname l@#1\endcsname
\fi
#2}}
\providecommand{\BIBdecl}{\relax}
\BIBdecl

\bibitem{computational_paralinguistics}
A.~Batliner and B.~Schuller, \emph{{Computational Paralinguistics: Emotion,
  Affect and Personality in Speech and Language Processing}}.\hskip 1em plus
  0.5em minus 0.4em\relax New York: John Wiley \& Sons, Incorporated, 2013.

\bibitem{batliner_recognition_of_emotions_in_speech}
A.~Batliner, B.~Schuller, D.~Seppi, S.~Steidl, L.~Devillers, L.~Vidrascu,
  T.~Vogt, V.~Aharonson, and N.~Amir, ``{The Automatic Recognition of Emotions
  in Speech},'' in \emph{Emotion-Oriented Systems}.\hskip 1em plus 0.5em minus
  0.4em\relax Berlin, Heidelberg: Springer Berlin Heidelberg, 2010, pp. 71--99.

\bibitem{eva_paperi}
E.~St{\aa}hlberg-Forsen, A.~Aija, B.~Kaasik, R.~Latva,
  S.~Ahlqvist-Bj{\"o}rkroth, L.~Toome, L.~Lehtonen, and S.~Stolt, ``{The
  validity of the Language Environment Analysis system in two neonatal
  intensive care units},'' \emph{Acta Paediatrica}, 2021.

\bibitem{schuller_cross_corpus_ser}
B.~{Schuller}, B.~{Vlasenko}, F.~{Eyben}, M.~{Wöllmer}, A.~{Stuhlsatz},
  A.~{Wendemuth}, and G.~{Rigoll}, ``{Cross-Corpus Acoustic Emotion
  Recognition: Variances and Strategies},'' \emph{IEEE Transactions on
  Affective Computing}, vol.~1, no.~2, pp. 119--131, 2010.

\bibitem{ser_f_similarity}
B.~Zhang, Y.~Kong, G.~Essl, and E.~M. Provost, ``{f-Similarity Preservation
  Loss for Soft Labels: A Demonstration on Cross-Corpus Speech Emotion
  Recognition},'' \emph{Proceedings of the AAAI Conference on Artificial
  Intelligence}, vol.~33, no.~01, pp. 5725--5732, 2019.

\bibitem{ser_semisupervised_autoencoders}
J.~{Deng}, X.~{Xu}, Z.~{Zhang}, S.~{Frühholz}, and B.~{Schuller},
  ``{Semisupervised Autoencoders for Speech Emotion Recognition},''
  \emph{IEEE/ACM Transactions on Audio, Speech, and Language Processing},
  vol.~26, no.~1, pp. 31--43, 2018.

\bibitem{ser_domain_adversarial}
M.~Abdelwahab and C.~Busso, ``{Domain Adversarial for Acoustic Emotion
  Recognition},'' \emph{IEEE/ACM Trans. Audio, Speech and Language Processing},
  vol.~26, no.~12, p. 2423–2435, 2018.

\bibitem{ser_da_cross_lingual}
H.~{Sagha}, J.~{Deng}, M.~{Gavryukova}, J.~{Han}, and B.~{Schuller}, ``{Cross
  lingual speech emotion recognition using canonical correlation analysis on
  principal component subspace},'' in \emph{Proc. ICASSP}, 2016, pp.
  5800--5804.

\bibitem{ser_universum_autoencoder}
J.~{Deng}, X.~{Xu}, Z.~{Zhang}, S.~{Frühholz}, and B.~{Schuller}, ``{Universum
  Autoencoder-Based Domain Adaptation for Speech Emotion Recognition},''
  \emph{IEEE Signal Processing Letters}, vol.~24, no.~4, pp. 500--504, 2017.

\bibitem{ser_cross_lingual}
S.~Latif, J.~Qadir, and M.~Bilal, ``{Unsupervised Adversarial Domain Adaptation
  for Cross-Lingual Speech Emotion Recognition},'' in \emph{2019 8th
  International Conference on Affective Computing and Intelligent Interaction
  (ACII)}, 2019, pp. 732--737.

\bibitem{ser_al_random_fields}
Z.~{Zhao} and X.~{Ma}, ``{Active Learning for Speech Emotion Recognition Using
  Conditional Random Fields},'' in \emph{2013 14th ACIS International
  Conference on Software Engineering, Artificial Intelligence, Networking and
  Parallel/Distributed Computing}, 2013, pp. 127--131.

\bibitem{ser_dnn_al}
M.~{Abdelwahab} and C.~{Busso}, ``{Active Learning for Speech Emotion
  Recognition Using Deep Neural Network},'' in \emph{2019 8th International
  Conference on Affective Computing and Intelligent Interaction (ACII)}, 2019,
  pp. 1--7.

\bibitem{ser_largescale_internet}
J.~{Jia}, S.~{Zhou}, Y.~{Yin}, B.~{Wu}, W.~{Chen}, F.~{Meng}, and Y.~{Wang},
  ``{Inferring Emotions From Large-Scale Internet Voice Data},'' \emph{IEEE
  Transactions on Multimedia}, vol.~21, no.~7, pp. 1853--1866, 2019.

\bibitem{lssed}
W.~Fan, X.~Xu, X.~Xing, W.~Chen, and D.~Huang, ``{LSSED: a large-scale dataset
  and benchmark for speech emotion recognition},'' \emph{arXiv preprint arXiv:
  2102.01754}, 2021.

\bibitem{tuni_mal}
Z.~{Shuyang}, T.~{Heittola}, and T.~{Virtanen}, ``{Active learning for sound
  event classification by clustering unlabeled data},'' in \emph{Proc. ICASSP},
  2017, pp. 751--755.

\bibitem{pearson_distance}
K.~A.~S. {Immink} and J.~H. {Weber}, ``{Minimum Pearson Distance Detection for
  Multilevel Channels With Gain and/or Offset Mismatch},'' \emph{IEEE
  Transactions on Information Theory}, vol.~60, no.~10, pp. 5966--5974, 2014.

\bibitem{k_medoids_algorithm}
H.-S. Park and C.-H. Jun, ``{A Simple and Fast Algorithm for K-Medoids
  Clustering},'' \emph{Expert Syst. Appl.}, vol.~36, no.~2, p. 3336–3341,
  2009.

\bibitem{tuni_adversarial_domain_adaptation}
K.~Drossos, P.~Magron, and T.~Virtanen, ``{Unsupervised Adversarial Domain
  Adaptation Based on The Wasserstein Distance For Acoustic Scene
  Classification},'' in \emph{2019 IEEE Workshop on Applications of Signal
  Processing to Audio and Acoustics (WASPAA)}, 2019, pp. 259--263.

\bibitem{wgan_original}
M.~Arjovsky, S.~Chintala, and L.~Bottou, ``{Wasserstein Generative Adversarial
  Networks},'' in \emph{Proc. ICML}, D.~Precup and Y.~W. Teh, Eds.,
  vol.~70.\hskip 1em plus 0.5em minus 0.4em\relax International Convention
  Centre, Sydney, Australia: PMLR, 2017, pp. 214--223.

\bibitem{LENAXu}
D.~Xu, U.~Yapanel, S.~Gray, J.~Gilkerson, J.~Richards, and J.~Hansen, ``{Signal
  processing for young child speech language development},'' in \emph{Proc. 1st
  Workshop on Child, Computer, and Interaction (WOCCI-2008)}, 2008.

\bibitem{siirila_kati_gradu}
K.~Siirilä, ``{Language Environment Analysis {(LENA)} -menetelmän
  validiteetti keskosvauvojen ääniympäristön arvioinnissa},'' Master's
  thesis, University of Turku, 2019.

\bibitem{lena_evaluation}
A.~Cristia, M.~Lavechin, C.~Scaff, M.~Soderstrom, C.~Rowland,
  O.~R{\"a}s{\"a}nen, J.~Bunce, and E.~Bergelson, ``{A thorough evaluation of
  the Language Environment Analysis {(LENA)} system},'' \emph{Behavior Research
  Methods}, 2020.

\bibitem{emodb}
F.~Burkhardt, A.~Paeschke, M.~Rolfes, W.~Sendlmeier, and B.~Weiss, ``{A
  database of German emotional speech},'' in \emph{9th European Conference on
  Speech Communication and Technology}, vol.~5, 2005, pp. 1517--1520.

\bibitem{enterface}
O.~{Martin}, I.~{Kotsia}, B.~{Macq}, and I.~{Pitas}, ``{The eNTERFACE' 05
  Audio-Visual Emotion Database},'' in \emph{22nd International Conference on
  Data Engineering Workshops (ICDEW'06)}, 2006, pp. 1--8.

\bibitem{finnish_corpus}
M.~Airas and P.~Alku, ``{Emotions in Vowel Segments of Continuous Speech:
  Analysis of the Glottal Flow Using the Normalised Amplitude Quotient},''
  \emph{Phonetica}, vol.~63, pp. 26--46, 2006.

\bibitem{einari_dippa}
E.~Vaaras, ``{Automatic Emotional Speech Analysis from Daylong Child-Centered
  Recordings from a Neonatal Intensive Care Unit},'' Master's thesis, Tampere
  University, 2021.

\bibitem{ravdess}
S.~R. Livingstone and F.~A. Russo, ``{The Ryerson Audio-Visual Database of
  Emotional Speech and Song (RAVDESS): A dynamic, multimodal set of facial and
  vocal expressions in North American English},'' \emph{PLoS ONE}, vol.~13,
  no.~5, pp. 1--35, 2018.

\bibitem{gemaps}
F.~{Eyben}, K.~R. {Scherer}, B.~W. {Schuller}, J.~{Sundberg}, E.~{André},
  C.~{Busso}, L.~Y. {Devillers}, J.~{Epps}, P.~{Laukka}, S.~S. {Narayanan}, and
  K.~P. {Truong}, ``{The Geneva Minimalistic Acoustic Parameter Set (GeMAPS)
  for Voice Research and Affective Computing},'' \emph{IEEE Transactions on
  Affective Computing}, vol.~7, no.~2, pp. 190--202, 2016.

\bibitem{opensmile}
F.~Eyben, F.~Weninger, F.~Gross, and B.~Schuller, ``{Recent developments in
  openSMILE, the Munich open-source multimedia feature extractor},'' in
  \emph{MM 2013 - Proceedings of the 2013 ACM Multimedia Conference}.\hskip 1em
  plus 0.5em minus 0.4em\relax ACM, 2013, pp. 835--838.

\bibitem{ser_unite_or_vote}
B.~Schuller, Z.~Zhang, F.~Weninger, and G.~Rigoll, ``{Using Multiple Databases
  for Training in Emotion Recognition: To Unite or to Vote?}'' in \emph{Proc.
  INTERSPEECH}, 2011, pp. 1553--1556.

\bibitem{ser_unsupervised_learning_cross_corpus}
Z.~{Zhang}, F.~{Weninger}, M.~{Wöllmer}, and B.~{Schuller}, ``{Unsupervised
  learning in cross-corpus acoustic emotion recognition},'' in \emph{2011 IEEE
  Workshop on Automatic Speech Recognition Understanding}, 2011, pp. 523--528.

\bibitem{elu_original}
D.-A. Clevert, T.~Unterthiner, and S.~Hochreiter, ``{Fast and Accurate Deep
  Network Learning by Exponential Linear Units (ELUs)},'' in \emph{4th
  International Conference on Learning Representations}, 2016.

\bibitem{adam_original}
D.~P. Kingma and J.~Ba, ``{Adam: A Method for Stochastic Optimization},'' in
  \emph{3rd International Conference on Learning Representations}, 2015.

\bibitem{leakyrelu_original}
A.~L. Maas, A.~Y. Hannun, and A.~Y. Ng, ``{Rectifier Nonlinearities Improve
  Neural Network Acoustic Models},'' in \emph{Proc. ICML}, 2013.

\bibitem{rmsprop_original}
T.~Tieleman and G.~Hinton, ``{Lecture 6.5---RMSProp: Divide the gradient by a
  running average of its recent magnitude},'' COURSERA: Neural Networks for
  Machine Learning, 2012.

\end{thebibliography}

\end{document}